\documentclass[final, numberedheadings ]{aipproc}

\usepackage{caption}
\usepackage{ amssymb }

\def\<{\left \langle}
\def\>{\right \rangle}
\def\[{\left\lbrack}
\def\]{\right\rbrack}
\def\({\left(}
\def\){\right)}
\newcommand{\be}{\begin{equation}}
\newcommand{\ee}{\end{equation}}
\newcommand{\ea}{\end{eqnarray}}
\newcommand{\ba}{\begin{eqnarray}}

\newcommand{\bs}{\begin{sideways}}
\newcommand{\es}{\end{sideways}}
 
\newcommand{\Newt}{{\mbox{\tiny Newt}}}

\def\beq{\begin{eqnarray}}
\def\eeq{\end{eqnarray}}
\def\ln{\,\mbox{ln}\,}

\newcommand{\mnras}{Mon. Not. Roy. Astron. Soc.}
\newcommand{\apj}{Astrophys. J.}
\newcommand{\aj}{Astron. J.}

\layoutstyle{6x9}

\begin{document}

\title{\textbf{Galaxy Rotation Curves from General Relativity with Infrared Renormalization Group Effects}} 

\author{Davi C. Rodrigues}{address={Departamento de F\'isica, CCE, Universidade Federal do Esp\'irito Santo, 29075-910, Vit\'oria, ES, Brazil.}}
\author{Patricio S. Letelier}{address={Departamento de Matem\'atica Aplicada, IMECC, Universidade Estadual de Campinas, 13083-859, Campinas, SP, Brazil.}}
\author{Ilya L. Shapiro}{address={Departamento de F\'isica, ICE, Universidade Federal de Juiz de Fora, 36036-330, Juiz de Fora, MG, Brazil.}}

\begin{abstract}
We review our contribution to infrared Renormalization Group (RG) effects to General Relativity in the context of galaxies. Considering the effective action approach to Quantum Field Theory in curved background, we argued that the proper RG energy scale, in the weak field limit, should be related to the Newtonian potential. In the galaxy context, even without dark matter, this led to a remarkably small  gravitational coupling $G$ variation (about or less than $ 10^{-12}$ of its value per light-year), while also capable of generating galaxy rotation curves about as good as the best phenomenological dark matter profiles (considering both the rotation curve shape and the expected mass-to-light ratios). Here we also comment on related developments, open issues and perspectives. 
\end{abstract}

\maketitle

\section{Introduction}
Currently there is a large body of data coming from cosmological and astrophysical observations that is mostly consistent with the existence of dark matter.  Such observations also suggest that the hypothesized particles that constitute dark matter have very small cross section and travel at speeds much lower than light. These lead to the cold dark matter framework, which is one of the pillars of the current standard cosmological model $\Lambda$CDM.

It is not only tempting, but mandatory to check if such dark matter particles exist (by detecting them in laboratory based experiments, for instance) and also to check if the gravitational effects that lead to the dark matter hypothesis could follow from a more detailed and complete approach to gravity. The effects of pure classical General Relativity at galaxies have been studied for a long time and, considering galaxy rotation curves, the differences between General Relativity and Newtonian gravity are negligible, since in a galaxy matter moves at speeds much lower than that of light and is typically subject to weak gravitational fields ($\Phi \ll c^2$), which leads to the Newtonian limit of General Relativity \footnote{There are some proposals that consider General Relativity  in the context of galaxies which do not lead to Newtonian gravity, see e.g. \cite{Cooperstock:2006dt, Vogt:2005hn, Cooperstock:2007sc, Vogt:2005va, Vogt:2005hv}.  It is not impossible that a reasonable explanation for galaxy rotation curves may rely on similar approaches, nevertheless up to now none of such proposals have found a baryonic mass distribution that is in conformity with astrophysical expectations.}. 

There is however a newer approach that may change considerably the role of dark matter, while following standard physical principles. Namely, the investigation of the running of the gravitational coupling parameter $G$ on large scales as induced by the renormalization group framework.

The running of coupling constants is a well known phenomenon within 
Quantum Field Theory. It is well-known that the renormalization group 
method can be extended to quantum field theory on curved space time
and to some models of quantum gravity (see, e.g., \cite{Buchbinder:1992rb}), such that  the beta functions can be interpreted in this framework. Concerning 
the high energy (UV) behavior, there is hope that the running of $G$ 
in quantum gravity may converge to a non-Gaussian fixed point 
(asymptotic safety) \cite{Niedermaier:2006wt,Weinberg:2009bg}. Our present concern is, however, not about 
the UV completeness, but with the behavior of $G$ in the far infrared 
regime (IR). In the electromagnetic case the IR behavior corresponds 
to the Appelquist-Carazzone decoupling \cite{Appelquist:1974tg} (see e.g., \cite{Goncalves:2009sk} for a recent derivation of this theorem). In the case of gravity the same effect of decoupling has been obtained in \cite{Gorbar:2002pw,Gorbar:2003yt},
but only for the higher derivative terms in the gravitational action. 
It remains unclear whether such decoupling takes place for the other terms. This possibility was studied on phenomenological grounds a number of times before, e.g. \cite{Shapiro:2004ch,Reuter:2004nx}.

In  \cite{Rodrigues:2009vf} we presented new results on the application of renormalization group (RG) corrections to General Relativity (GR) in the astrophysical domain. Previous attempts to apply this picture to galaxies have considered for simplicity point-like galaxies. We extended previous considerations by identifying the proper renormalization group energy scale $\mu$ and by evaluating the consequences considering the observational data of disk galaxies. Also we propose a natural choice for the identification of $\mu$, linking it to the local value of the Newtonian gravitational potential. With this choice, the renormalization group-based approach is capable to mimic dark matter effects with great precision. This picture induces a very small variation on the gravitational coupling parameter $G$, namely a variation of about $10^{-7}$ of its value across $10^5$ light-years. We call our model RGGR, in reference to renormalization group effects in General Relativity.

In order to evaluate the observational consequences of the RGGR model and to compare it to other proposals, 
recent high quality observational data \cite{2008AJ....136.2648D,Gentile:2004tb} from nine regular spiral galaxies were mass-modelled using the standard procedures for the baryonic part, and four different models for the ``dark'' component: i) the RGGR model; ii) one of the most phenomenological successful dark matter profiles, the Isothermal profile \cite{1991MNRAS.249..523B}; and two alternative models which were built to avoid the need for the dark matter: iii) the Modified Newtonian Dynamics (MOND) \cite{1983ApJ...270..365M,1983ApJ...270..371M} and iv) the Scalar-Tensor-Vector Gravity (STVG) \cite{Moffat:2005si}. The latter is a recent relativistic proposal that is capable of dealing with galaxy rotation curves and other phenomena usually attributed to dark matter. For galaxy rotation curves phenomena, STVG becomes equivalent to a similar proposal called MSTG \cite{Moffat:2004bm, Brownstein:2005zz, Brownstein:2009gz}. The model parameters that we use to fit galaxies in the STVG framework can be found in ref. \cite{Brownstein:2005zz}. While for MOND we use the $a_0$ value as given in \cite{Sanders:2002pf}.

The quality of the rotation curve fits and total stellar mass as inferred from the RGGR model is perfectly satisfactory considering both the general behavior of the model and its results when applied to nine particular galaxies, as analyzed in \cite{Rodrigues:2009vf}. It is about the same of the Isothermal profile quality, while it seems significantly better than the quality of MOND and STVG. In the case of MOND, we did numerical experiences with $a_0$ as a free parameter and found that, albeit the concordance with the shape of the rotation curve can considerably increase in this case, the concordance with the expected stellar mass-to-light ratios remains unsatisfactory (similar conclusions have also appeared in some recent papers, e.g. \cite{Gentile:2004tb,Gentile:2010xt,Famaey:2005fd}, and it seems that the concordance can only be improved by adjusting the MOND's $\mu(x)$ function in an {\it ad-hoc} way).

\section{The running of $G$}

The $\beta$-function for  the gravitational coupling parameter $G$ has been discussed in the framework of different approaches to Quantum Gravity and Quantum Field Theory in curved space-time. In \cite{Rodrigues:2009vf} we followed the derivation used previously in \cite{Shapiro:2004ch}.  If  $G$ does not behave as  a constant  in the far  IR limit, it was argued in \cite{Shapiro:2004ch} (and recently in more details in \cite{Farina:2011me}) that the logarithmic running of $G$  is a direct consequence of covariance and must hold in all loop orders.  As far as direct derivation of the physical running of $G$ is not available, it is worthwhile to explore the possibility of a logarithmically running $G$ at the phenomenological level.

Consider the following infrared $\beta$-function for General Relativity,
\be
	 \beta_{G^{-1}} \equiv \mu \frac{dG^{-1}}{d \mu} = 2 \nu \,  \frac{M_{\mbox{\tiny Planck}}^{2}}{c \, \hbar} = 2 \nu G_0^{-1}.
	\label{betaG}
\ee

Equation (\ref{betaG}) leads to the logarithmically  varying $G(\mu)$ function,
\be
	\label{gmu}
	G(\mu) = \frac {G_0}{ 1 + \nu \ln(\mu^2/\mu_0^2)},
\ee
where $\mu_0$ is a reference scale introduced such that $G(\mu_0) =G_0 $. The constant $G_0$ is the gravitational constant as measured in the Solar System  (actually, there is no need to be very precise on where $G$ assumes the value of $G_0$, due to the smallness of the variation of $G$). The dimensionless constant $\nu$ is a phenomenological parameter which depends on the details of the quantum theory leading to eq. (\ref{gmu}). Since we have no means to compute the latter from first principles, its value should be fixed from observations. It will be shown that even a very small $\nu$ can lead to observational consequences at galactic scales.

The action for this model is simply the Einstein-Hilbert one in which $G$ appears inside the integral, namely,
\be
	S_{\mbox{\tiny RGGR}}[g] = \frac {c^3}{16 \pi }\int \frac {R  } G \, \sqrt{-g} \,  d^4x.
	\label{rggraction}
\ee
In the above, $G$ should be understood as an external scalar field that satisfies (\ref{gmu}).  Since for the problem of galaxy rotation curves  the cosmological constant effects are negligible, we have not written the $\Lambda$ term above. Nevertheless, for a complete cosmological picture, $\Lambda$ is necessary and it also runs covariantly with the RG flow of $G$ (see e.g.,\cite{Shapiro:2004ch}). 

There is a simple procedure to map solutions from the Einstein equations with the gravitational constant $G_0$  into RGGR solutions. One need not to follow this route, one may find all the dynamics from the RGGR equations of motion, which can be found by a direct variation of the action (\ref{rggraction}) in respect to the metric, leading to equations of motion that have the same form of those of a scalar-tensor gravity\footnote{We stress that it is only the from since RGGR is not a type of scalar-tensor gravity, and $G$ is not a fundamental field of the model.}. In this review, we will proceed to find RGGR solutions via a conformal transformation of the Einstein-Hilbert action, and to this end first we write 
\be
	G = G_0 + \delta G,
\ee 
and we assume $\delta G / G_0 \ll 1$, which will be justified latter. Introducing the conformally related metric
\be
	\bar g_{\mu \nu} \equiv \frac {G_0}{G} g_{\mu \nu}, 
	\label{ct}
\ee
the RGGR action can be written as
\be
	S_{\mbox{\tiny RGGR}}[g] = S_{\mbox{\tiny EH}}[\bar g] + O(\delta G^2),
\ee
where $S_{\mbox{\tiny EH}}$ is the Einstein-Hilbert action with $G_0$ as the gravitational constant. The above suggest that the RGGR solutions can be generated from the Einstein equations solutions via the conformal transformation (\ref{ct}). Indeed, within a good approximation, one can check that this relation persists when comparing the RGGR equations of motion to the Einstein equations even in the presence of matter \cite{Rodrigues:2009vf}.

In the context of rotation curves of galaxies, standard General Relativity gives essentially the same predictions of Newtonian gravity. The Newtonian potential $ \Phi_\Newt$ is related to the metric by 
\be
	\bar g_{00} = - \left ( 1  + \frac {2 \Phi_\Newt}{c^2} \right ).
\ee
Hence, using eq. (\ref{ct}), the effective RGGR potential $\Phi$ in the non-relativistic limit is given by
\be
	\Phi = \Phi_\Newt + \frac {c^2}2 \frac{\delta G}{G_0}.
\ee
An equivalent result can also be found from the evaluation of a test particle geodesics \cite{Rodrigues:2009vf}. In the context of weak gravitational fields $\Phi_\Newt/ c^2 \ll 1$ (with $\Phi_\Newt = 0$ at spatial infinity) holds, and hence the term $\delta G/G_0$ should not be neglected.

In order to derive a test particle acceleration, we have to specify the proper energy scale $\mu$ for the problem setting in question, which is a time-independent gravitational phenomena in the weak field limit. This is a recent area of exploration of the renormalization group application, where the usual procedures for high energy scattering of particles cannot be applied straightforwardly. Previously to \cite{Rodrigues:2009vf} the selection of $\mu \propto 1/r$, where $r$ is the distance from a massive point, was repeatedly used, e.g. \cite{Reuter:2004nv,Dalvit:1994gf,Bertolami:1993mh,Goldman:1992qs, Shapiro:2004ch}. This identification adds a constant velocity proportional to $\nu$ to any rotation curve. Although it was pointed as an advantage due to the generation of ``flat rotation curves'' for galaxies, it introduced difficulties with the Tully-Fisher law, the Newtonian limit, and the behavior of the galaxy rotation curve close to the galactic center, since there the behavior is closer to the expected one without dark matter. In \cite{Rodrigues:2009vf} we introduced a $\mu$ identification that seems better justified both from the theoretical and observational points of view. The characteristic weak-field gravitational energy does not comes from the geometric scaling $1/r$, but from the Newtonian potential $\Phi_\Newt$. However, the straight relation $ \mu \propto \Phi_\Newt$ leads to $\mu \propto 1/r$ in the large $r$ limit; which is unsatisfactory on observational grounds (bad Newtonian limit and correspondence to the Tully-Fisher law). One way to recover the Newtonian limit is to impose a suitable cut-off, but this does not solves the Tully-Fisher issues \cite{Shapiro:2004ch}. Another one is to use \cite{Rodrigues:2009vf}
\be
	\frac {\mu}{\mu_0} =\left( \frac{\Phi_\Newt}{\Phi_0} \right)^\alpha,
	\label{muphi}
\ee
where $\Phi_0$ and $\alpha$ are constants. Apart from the condition $0 < \Phi_0 < c^2$ (i.e., essentially $\Phi_0$ is a reference Newtonian potential), the precise value of $\Phi_0$ is largely irrelevant for the problem of rotation curves. The relevant parameter is $\alpha$. It is a phenomenological parameter that depends on the mass of the system, and it must go to zero when the mass of the system goes to zero. This is necessary to have a good Newtonian limit. From the Tully-Fisher law, it is expected to increase monotonically with the increase of the mass. Such behavior is indeed found from the galaxy fits done in \cite{Rodrigues:2009vf}. In a recent paper, an upper bound on $\nu \alpha$ in the Solar System was derived \cite{Farina:2011me}. In galaxy systems, $\nu \alpha|_{\mbox{\tiny Galaxy}} \sim 10^{-7}$, while for the Solar System, whose mass is about $10^{-10}$ of that of a galaxy,  $\nu \alpha|_{\mbox{\tiny Solar System}} \lesssim 10^{-17}$. It shows that a linear decrease on $ \alpha$ with the mass is sufficient to satisfy both the current upper bound from the Solar System and the results from galaxies.

We also point that the above energy scale setting (\ref{muphi}) was recently re-obtained from a more theoretical perspective \cite{Domazet:2010bk}.

Once the $\mu$ identification is set, it is straightforward to find the rotation velocity for a static gravitational system sustained by its centripetal acceleration,
\be
	V^2_{\mbox{\tiny RGGR}} \approx V^2_\Newt \left ( 1 - \frac {\nu \, \alpha  \, c^2} {\Phi_\Newt} \right ).
	\label{v2rggr}
\ee

Contrary to Newtonian gravity, the value of the Newtonian 
potential at a given point does play a significant role 
in this approach. This sounds odd from the perspective of 
Newtonian gravity, but this is not so 
from the General Relativity viewpoint, since the latter has no 
free zero point of energy. In particular, the Schwarzschild 
solution is not invariant under a constant shift of the 
potential.

In the following, we will comment on the effect of the relation (\ref{v2rggr}) to galaxy rotation curves. First from a more general perspective, and then to the modeling of individual galaxies.

\section{Galaxy rotation curves}

Before proceeding to specific galaxy rotation curves modeling, it is more instructive to analyze general features of the relation (\ref{v2rggr}), and to compare it to the standard approach. In \cite{Rodrigues:2009vf} we analyze some general aspects and scaling laws of the RGGR  model, with no dark matter, in comparison to the isothermal profile; both of them, at this step, without gas and with an exponential stellar disk. In particular, it was pointed that the RGGR rotation curves have a reasonable shape to fit galaxies (i.e., no clear problems like increasing or decreasing too fast, oscillations...),  and that they effectively behave similarly to cored dark matter profiles at inner radii, whose effective core radius scales with the galaxy disk scale length. Further details in our paper.

\begin{figure}[ht]
	  \includegraphics[width=150mm]{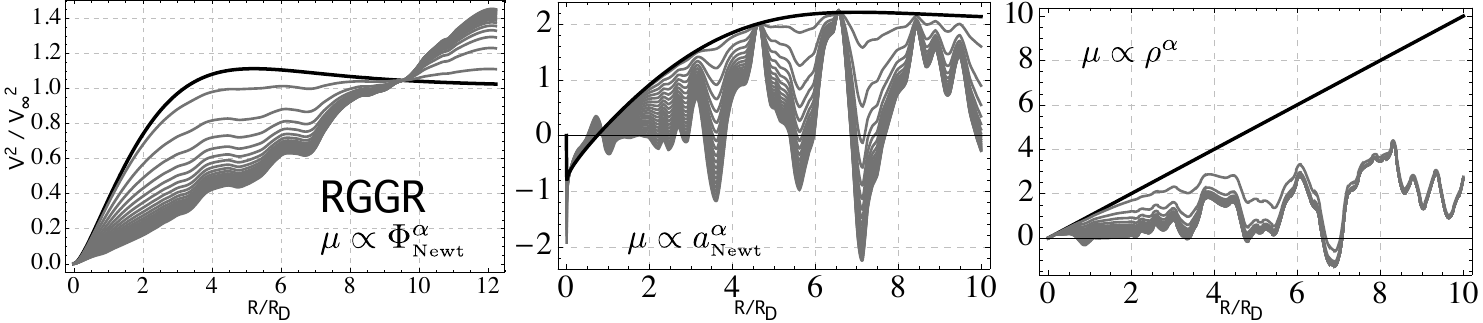}
\caption{The additional circular velocity squared induced by different  choices of $\mu$. The first plot refers to RGGR, the two others to different identifications of $\mu$: one depends on the Newtonian acceleration (a variation inspired on MOND) and the other on the (baryonic) matter density. All the plots above display the additional squared velocity of each model divided by $V^2_\infty \equiv \nu \alpha c^2$ and  as a function of $R/R_D$, where  $R$ is the the radial cylindrical coordinate and  $R_D$ is the stellar disk scale length. Black lines depict the additional velocity due to a pure exponential stellar disk, while the gray solid lines  take into account the gas mass $M_{\mbox{gas}}$ for different values of $f \equiv M_{\mbox{gas}}/M_{\mbox{stars}}$, with $f = 0.2, 0.7, 1.2,..., 9.7$ (i.e., the black lines stand for $f=0$). See \cite{Rodrigues:2009vf} for further details.}
\label{mus}
\end{figure}

We have also extended the previous analysis by adding a gas-like contribution (a re-scaled version of the NGC 3198 gaseous part). In particular, this numerically evaluates how the model behaves on the presence of density perturbations at large radii.  In the first plot of fig. (\ref{mus}) it is displayed the result for RGGR, which is remarkably good (a similar plot can be found in our original paper), while in the others plots in fig. (\ref{mus}) (presented at the Conference, but not in \cite{Rodrigues:2009vf}) one sees the results for the same mass distribution but different choices for the energy scale\footnote{These other choices are also unsatisfactory from the theoretical perspective, since they have no direct relation to the local energy of the gravitational field in the weak field regime ($\Phi_\Newt \ll c^2 $). } $\mu$.

From fig. (\ref{mus}), the two other proposals different from RGGR are seen to be unsuited as replacements for dark matter. In particular, both are too sensitive to the gas presence, and both eventually add negative contributions to the total circular velocity at large radii.

\begin{figure}[ht]
	  \includegraphics[width=100mm]{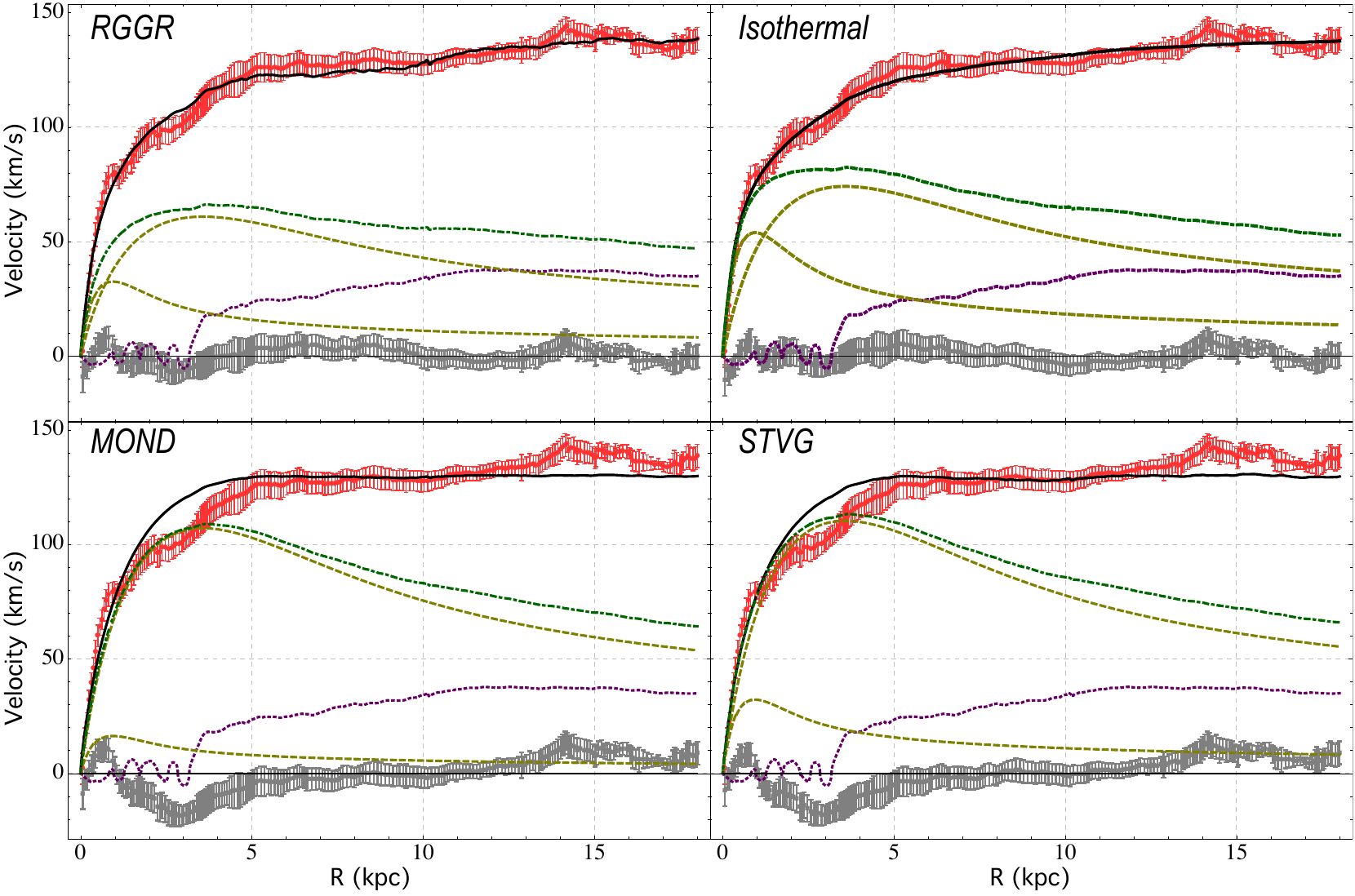}
\caption{NGC 2403 rotation curve fits. The red dots and its error bars are the rotation curve observational data, the gray ones close to the abscissa are the residues of the fit. The solid black line for each model is its best fit rotation curve, the dashed yellow curves are the stellar rotation curves from the bulge and disk components,  the dotted purple curve is the gas rotation curve, and the dot-dashed green curve is the resulting Newtonian, with no dark matter, rotation curve. }
\label{ngc2403}
\end{figure}

\begin{figure}[t]
	  \includegraphics[width=100mm]{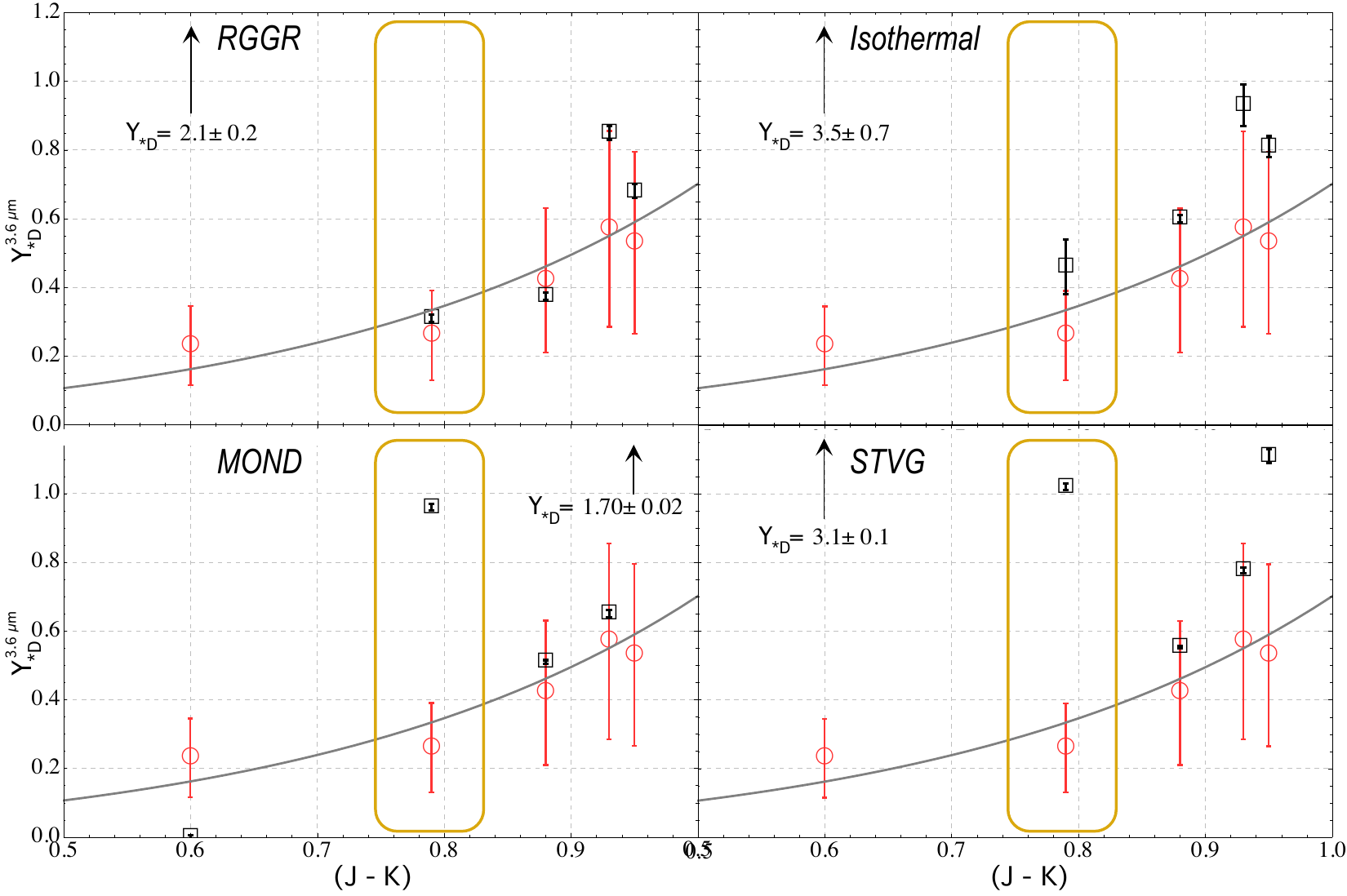}
\caption{Stellar disk mass-to-light ratio ($Y_{*D}$) in the $3.6  \mu m$ band  as a function of the color $J - K$.  Each galactic disk is represented above by an open circle, with a {\it reference} error bar of 50$\%$ of the $Y_{*D}$ value. The black open squares display the $Y_{*D}$ values and their associated 1$\sigma$ errors for each galaxy as inferred from the rotation curve fits for each model. The highlighted square and circle correspond to the NGC 2403 galactic disk mass-to-light ratios. See \cite{Rodrigues:2009vf} for further details.}
\label{masstolightplot}
\end{figure}

In \cite{Rodrigues:2009vf} we used a sample of nine high quality and regular rotation curves of disk galaxies from \cite{2008AJ....136.2648D, Gentile:2004tb}. In figs. (\ref{ngc2403}, \ref{masstolightplot}) we show one of ours results (see \cite{Rodrigues:2009vf} for the complete set and further details) in comparison to the results of three other models: a cored dark matter profile (Isothermal profile), the  Modified Newtonian Dynamics (MOND) and the recently proposed Scalar-Tensor-Vector Gravity (STVG). 

Due to the considerably large uncertainty in the total stellar mass of each stellar component (disk and bulge), we first use the total stellar mass as a free parameter for the fittings (achieved from a $\chi^2$ minimization considering the errors). At a second stage, we compare the resulting value with stellar population expectations, following the  standard approach.

On the free parameters of each model, we remark that besides the total stellar mass, the Isothermal profile has two additional free parameters, the RGGR model has a single free parameter ($\alpha$) while MOND and STVG have no free parameters that can vary from galaxy to galaxy. On the other hand, both of the latter depend on constants whose values are calibrated considering its best fit in a large sample of galaxies. We remark that the $\nu$ parameter in RGGR cannot vary from galaxy to galaxy, but $\alpha $ can, and galaxy rotation curves are sensible to the combination $\nu \alpha$, whose value is about the order of $10^{-7}$. The best fit for NGC 2403 yields $ \nu \alpha = (1.66 \pm0.01) \times 10^{-7}$.

\section{Conclusions}

We  presented a model, motivated by renormalization group corrections to the Einstein-Hilbert action, that introduces small inhomogeneities in the gravitational coupling across a galaxy (of about 1 part in $10^7$) and can generate galaxy rotation curves in agreement with the observational data, without the introduction of dark matter as a new kind of matter. Both High and Low Surface Brightness galaxies were tested \cite{Rodrigues:2009vf} . Considering the samples of  galaxies evaluated in \cite{Rodrigues:2009vf}, the quality of the RGGR rotation curves,  together with the corresponding mass-to-light ratios, is about the same than  the Isothermal profile quality, but with one less free parameter. We expect that similar results would hold in regard to other cored dark matter profiles, while our results seem better than those achieved by the NFW profile \cite{Rodrigues:2009vf}. We also compared the results of our model with MOND and STVG, and at face value our model yielded clearly better results.   

Our results can be seen as a next step compared to the 
previous models motivated by renormalization group effects in 
gravity, e.g. \cite{Shapiro:2004ch, Reuter:2007de}. Their original analyses 
could only yield a rough estimate on the galaxy rotation curves, 
since they were restricted to modeling a galaxy as a single 
point. Trying to extend this approach to real galaxies, we 
have shown that the proper scale 
for the renormalization group phenomenology is not of a 
geometric type, like the inverse of the distance, but is
related to the Newtonian potential with null boundary 
condition at infinity.

The essential feature for the RGGR rotation curves fits is the formula (\ref{v2rggr}), which is by itself a  simple formula that provides a very efficient description of galaxy rotation curves. 

There are several tests and implications of this model yet to be evaluated. In particular we are working on applying the RGGR framework to a larger sample of galaxies (including elliptical galaxies) \cite{FabrisGalaxies} and galaxy-galaxy strong lensing \cite{rodrigueslentes}. Related work on CMB, BAO and LSS in search for a new cosmological concordance model is also a work in progress \cite{toribioCMB}.

\bigskip

\noindent
{ \it \bf Acknowledgements}

\noindent
DCR thanks FAPESP and PRPPG-UFES for partial financial support. PSL thanks CNPq and FAPESP for partial financial support. The work of I.Sh. was partially supported by CNPq, FAPEMIG, FAPES and ICTP.

\end{document}